\documentclass{elsart}

\usepackage{epsfig}
\usepackage{amssymb}
\usepackage{graphicx}
\usepackage{dcolumn}
\usepackage{bm}
\newcommand{\be}{\begin{equation}}
\newcommand{\ee}{\end{equation}}
\newcommand{\bea}{\begin{eqnarray}}
\newcommand{\eea}{\end{eqnarray}}

\newcommand{\de}{\delta}

\newcommand{\om}{\omega}

\newcommand{\vq}{\vec q}

\newcommand{\mn}{\mu\nu}

\newcommand{\omp}{\om_\pi}
\newcommand{\omh}{\om_h}
\begin{document} 

\begin{frontmatter} 

\title{The $\rho$ meson in hot hadron matter and low mass dilepton spectra}

\author{Sabyasachi Ghosh, Sourav Sarkar and Jan-e Alam}
\medskip
\address{Variable Energy Cyclotron Centre, 1/AF, Bidhan Nagar, 
Kolkata - 700064,INDIA}

\begin{abstract}
The structure of the one loop self-energy graphs of the $\rho$ meson is
analyzed in the real time formulation of thermal field theory. The modified
spectral function of the $\rho$ meson in hot hadronic matter leads to
a large enhancement of lepton pair production below the bare peak of the $\rho$.
It has been shown that the effective temperature extracted from the inverse 
slope of the transverse momentum distributions for various invariant mass ($M$) 
windows of the pair can be used as an efficient tool to characterize different 
phases of the evolving matter.  

\end{abstract}


\end{frontmatter} 


We have studied dilepton production  
in relativistic heavy ion collision (HIC) experiments
with an aim to establish an observable distinction between emissions
from the QGP and hadronic phases by utilizing two kinds of medium effects.
One originates dynamically due to decay of vector mesons and their scattering 
from in-medium hadrons and the other through collective behaviour
of the fireball resulting in flow.
The emission rate of low $M$ lepton pairs is given
by the in-medium spectral function of the low mass vector mesons and is expressed
as \cite{Mclerran,GhoshJPG}
\be
\frac{dN}{d^4qd^4x}=\frac{\alpha^2}{\pi^3q^2}f_{BE}(q_0) 
\sum_{V=\rho,\omega,\phi}F_V^2m_V^2A_V(q_0,\vq)
\label{eq:dilrate2}
\ee 
The corresponding rate for emission from QGP is taken from \cite{JC}.
The spectral function of the $\rho$ meson which is known to play the dominant
role is given by
\bea
A_\rho&=&-\frac{1}{3}\left[\frac{2\sum{\rm
Im}\Pi^R_t}{(q^2-m_\rho^2-\sum\mathrm{Re}\Pi^R_t)^2
+(\sum{\rm Im}\Pi^{R}_t)^2}+{\rm \ long.\ comp.\ }\right]
\label{eq:spdef}
\eea
the sum running over the loop graphs. Here we consider one-loop diagrams
containing a pion and another meson $h$ ($=\pi$, $\omega$, $h_1$ and $a_1$).
The (retarded) self energy which appears in the spectral function
can be obtained in the real time formalism 
of thermal field theory from the 11-component of the self-energy which
is a 2$\times 2$ matrix in this approach~\cite{Bellac,MallikRT}. One thus writes for the
$\pi-h$ loop,
\be
\Pi_{\mn}^{11}(q)=i\int\frac{d^4k}{(2\pi)^4}N_{\mn}(q,k)D_\pi ^{11}(k)D_h^{11}(q-k)
\ee
where $D^{11}(q)$ represents thermal propagators and the factor $N_{\mn}$ 
includes tensor 
structures associated with the two vertices and those of the vector propagator
in the loop. From the trace and the 00-component, one obtains the transverse and
the longitudinal polarisations. The imaginary part in the present case basically 
decides the shape and magnitude of the spectral function, the real part 
causing a minor effect~\cite{GhoshJPG,Ghosh1}. Confining to positive values of $q^2$ and
$q_0$ we have,
\bea
&&{\rm Im}\Pi_{t,l}^R(q_0,\vq)=-\pi\int\frac{d^3\vec k}{(2\pi)^3 4\omp\omh}
\left[N_{t,l}(k_0=\omp)\{(1+n(\omp)+n(\omh))\right.\nonumber\\
&&\de(q_0-\omp-\omh)\}\left.+N_{t,l}(k_0=-\omp)\{(n(\omp)-n(\omh))\de(q_0+\omp-\omh)\}\right]~.
\label{eq:impi}
\eea
The first term is non-vanishing for $q^2\ge (m_h+m_{\pi})^2$
producing the unitary cut and
the second is non-vanishing for $q^2\ge (m_h-m_{\pi})^2$ giving 
the Landau cut.

\begin{figure}
\includegraphics[scale=0.29]{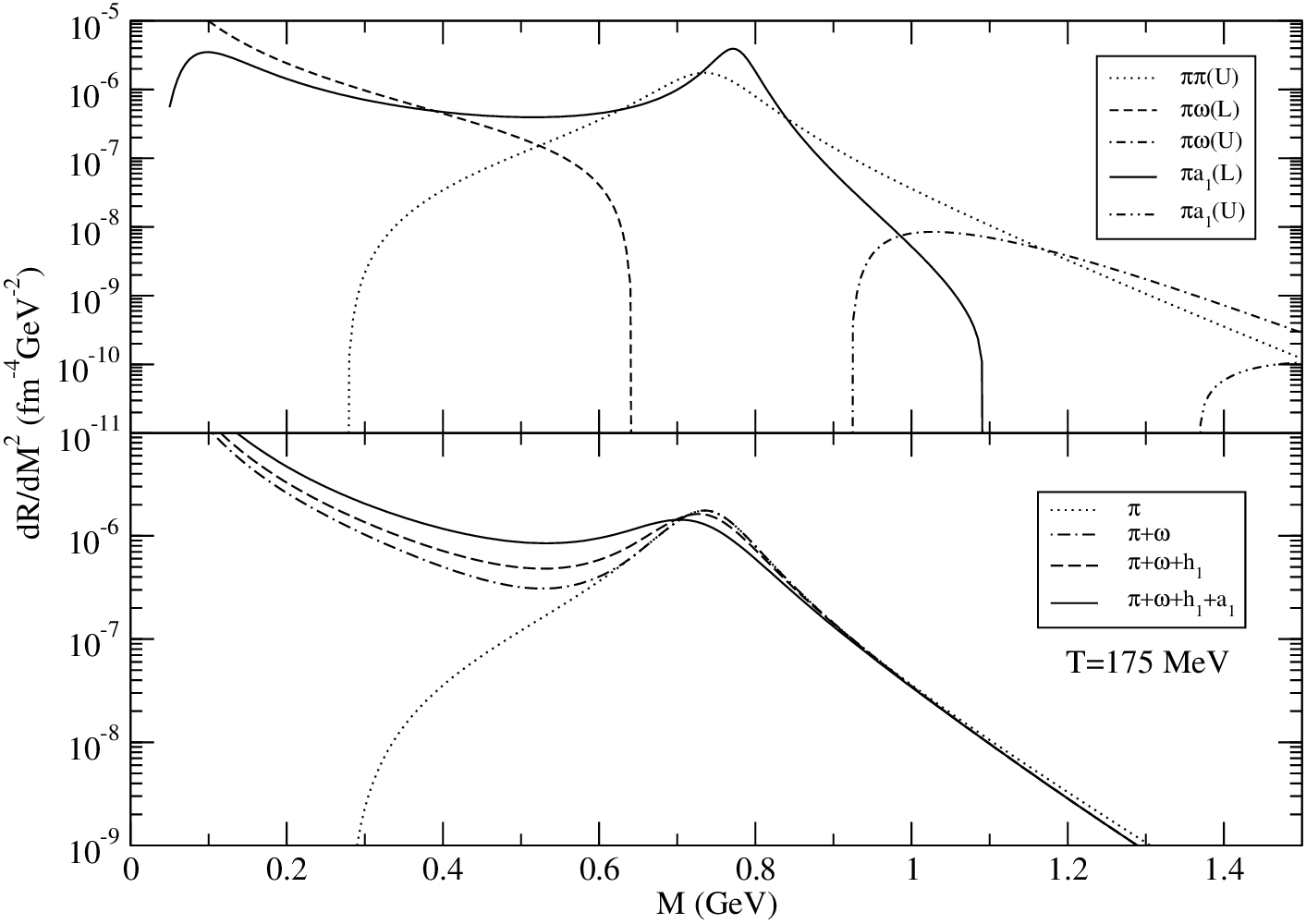}
\includegraphics[scale=0.29]{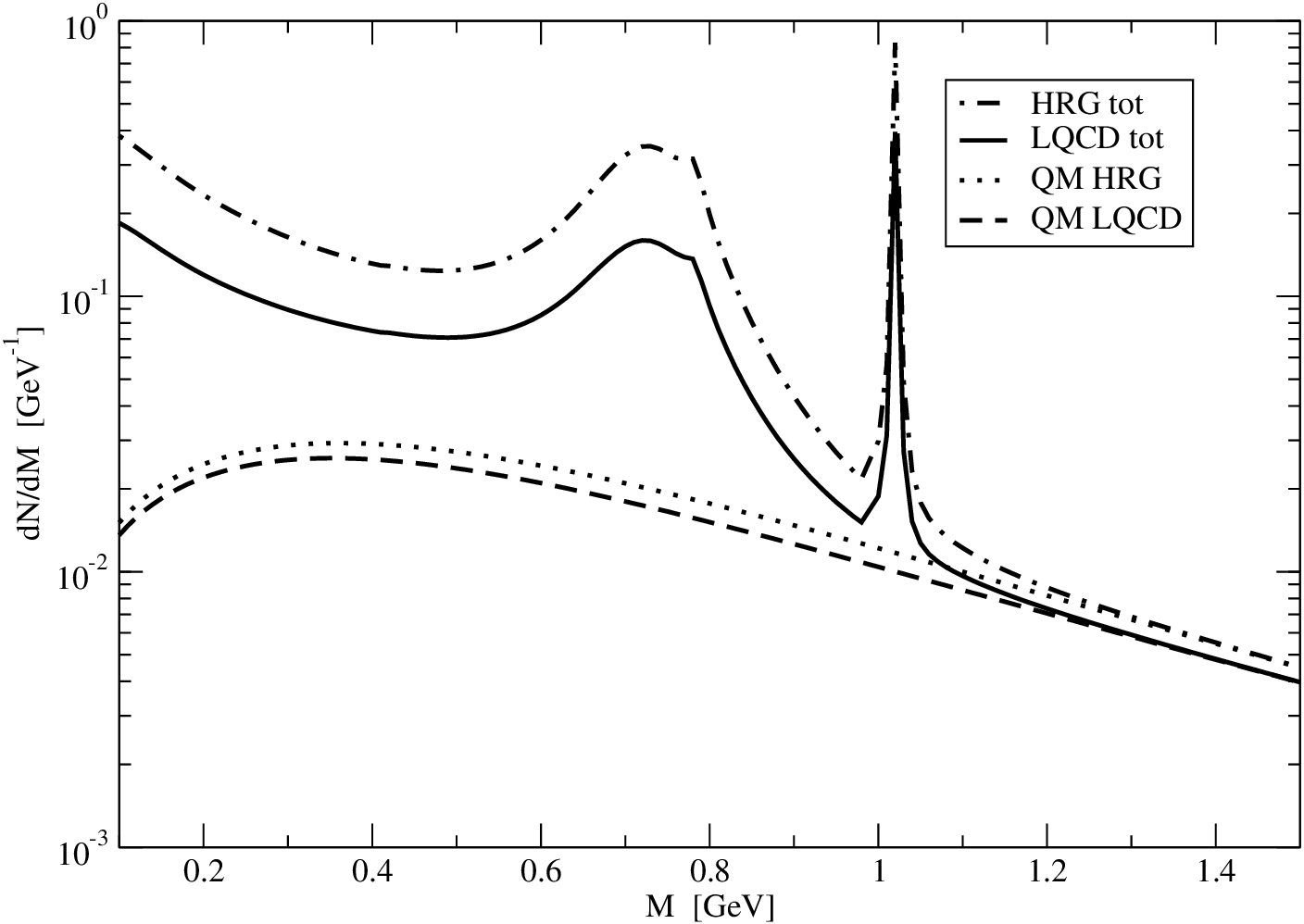}
\caption{Left panel (upper compartment) shows contributions from the 
discontinuities of the  self-energy graphs
to the dilepton emission rate at $T=175$ MeV. $L$ and $U$ denote the
Landau and unitary cut contribution. Left panel (lower compartment)
shows contributions from
the different mesons in the loop.
Right panel displays invariant mass distribution of dileptons from 
hadronic matter (HM)  and from (QM) at LHC energies.}
\label{fig:dil_LU}
\end{figure}

We plot in Fig.~\ref{fig:dil_LU} (left panel, upper compartment) 
the dilepton emission rate keeping only the 
$\rho$ contribution in Eq.~\ref{eq:dilrate2} in which we show the 
relative contributions from
the cuts in the $\pi-h$ loops keeping only one of them at a time. 
The unitary and Landau cuts for
the $ \pi,\omega,h_1 $ and $a_1$ are seen to contribute with different magnitudes
for different values of invariant mass ($M$) of $\rho$.
 The $\pi-\pi$ loop
has only the unitary cut and this contributes most significantly
to dilepton emission near the $\rho$ pole. 
For $\pi-\omega$ loop Landau cut ends at $M=m_\om-m_\pi$ and
the unitary cut starts at $M=m_\om+m_\pi$ , so there is no contribution at the $\rho$ pole.
 The Landau cut for the $\pi-a_1$ self-energy extends up to about 1100 MeV and
makes a substantial contribution both at and below the $\rho$ pole. The unitary cut
starts at a much higher value of $M$ and hence does not make a significant contribution
to the $\rho$ spectral function. 
In  Fig.~\ref{fig:dil_LU} (left panel, lower compartment) 
we show the cumulative contribution to the lepton pair yield
from the $\pi-\pi$ and the $\pi-h$ loops
in the region below the bare $\rho$ pole. 

We assume that an equilibrated QGP is formed in the HIC which
cools due to expansion and consequently reverts back to hadrons at 
$T_c=175$ MeV.  After the completion of the phase transition the 
hadronic matter cools further 
and eventually freezes out
first chemically  at a temperature $T_{ch}$ (=175 MeV) and then kinetically
at a temperature $T_F$ (=120 MeV). Relativistic hydrodynamics with cylindrical
symmetry~\cite{hvg} and boost invariance~\cite{jdb} along longitudinal
directions has been used for space-time description. 
Equation of states from lattice QCD and hadronic resonance 
gas have been used here(see~\cite{GhoshJPG} for details). 
The initial thermalization time, $\tau_i$  and initial temperature, $T_i$
are constrained to the total hadronic
multiplicity, $dN/dy\sim T_i^3\tau_i$. For $dN/dy=2600$,  $T_i=756$ MeV
and $\tau_i=0.1$ fm/c~\cite{GhoshJPG} for LHC. 
With these inputs the
space time integration is performed to obtain the transverse mass, 
$M_T (=\sqrt{M_{av}^2+p_T^2})$
spectra for various $M$ range - $M_{min}$ to $M_{max}$
with $M_{av}=(M_{max}+M_{min})/2$.  

\begin{figure}
\includegraphics[scale=0.285]{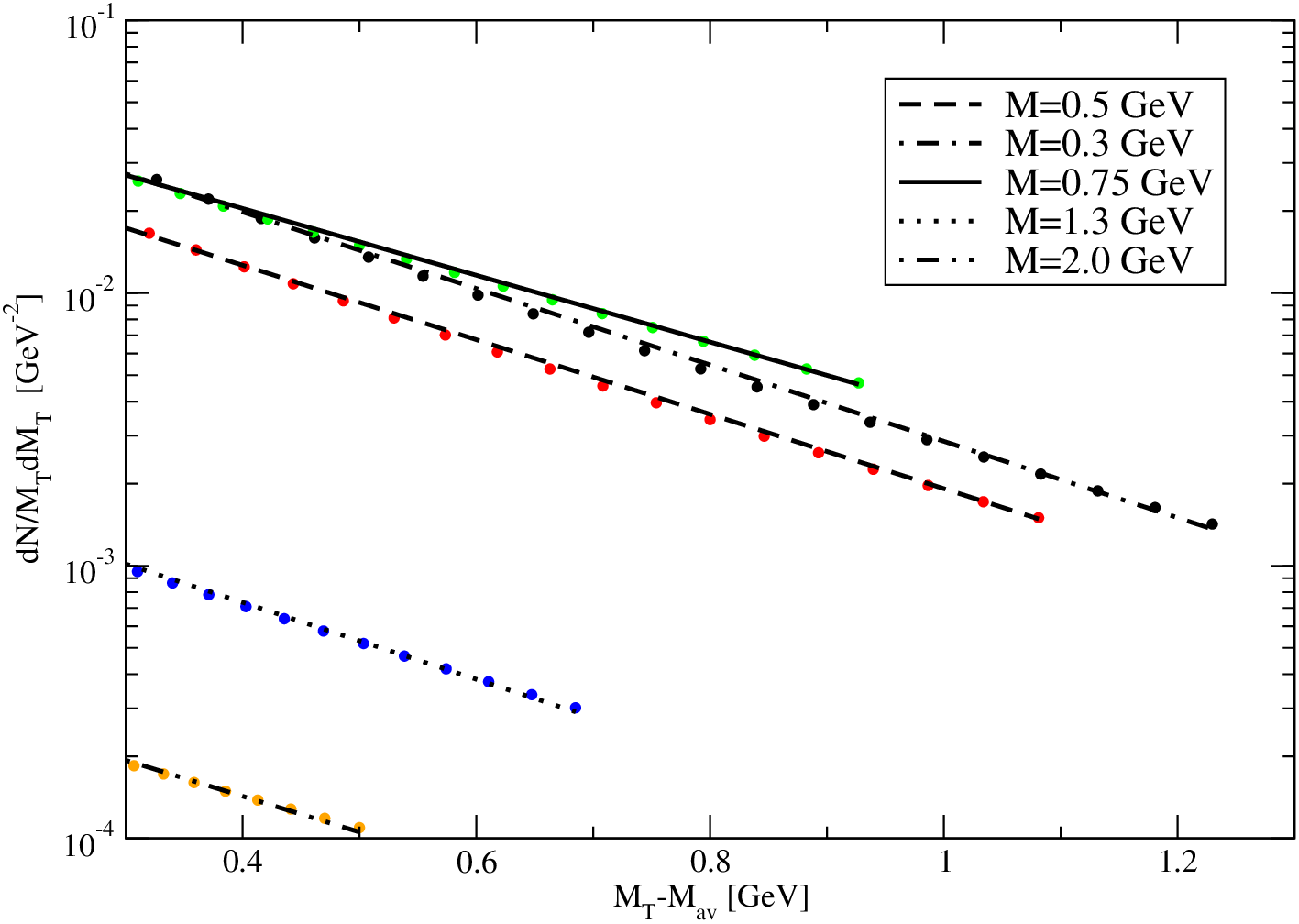}
\includegraphics[scale=0.285]{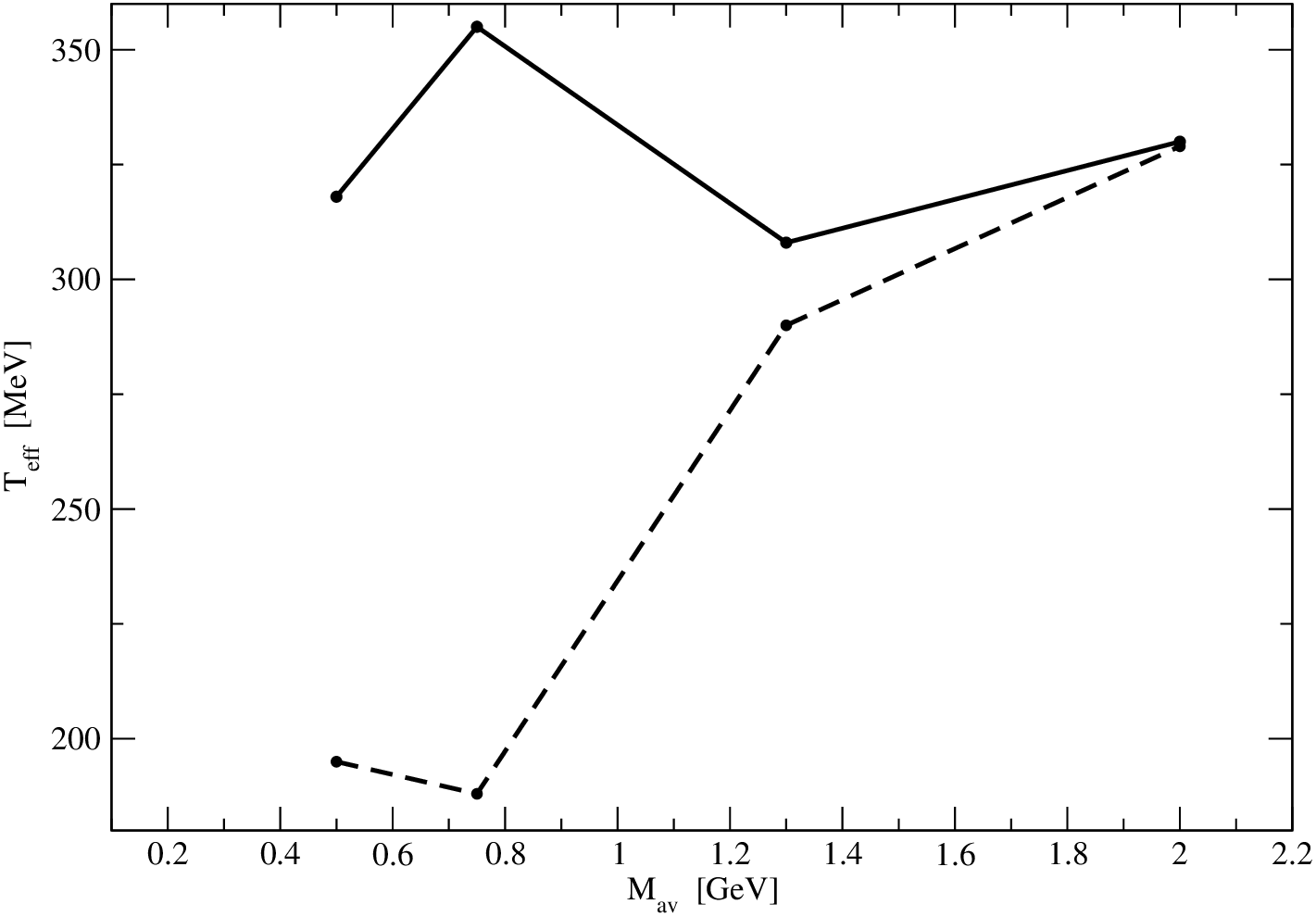}
\caption{Left panel shows variation of dilepton yield with $M_T-M_{av}$.
Right panel displays
the variation of the inverse slopes of $M_T$ distribution for 
different $M$-bins.
The dashed line is obtained by setting vanishing radial flow.}
\label{fig2}
\end{figure}

The  $M_T$ spectra for lepton pairs 
and their respective slopes for different mass windows  have been depicted 
in Fig.~\ref{fig2} in the left and right panels respectively.
It is evident from Fig.~\ref{fig:dil_LU} (right panel) that
the large $M$ pairs originate from early (QGP) phase where the
radial flow is minimal and the
low $M$ pairs  stem from the QGP as well as from the late (hadronic) 
phase containing reasonable amount of radial flow. However, 
pairs with $M\sim m_\rho$ are overwhelmingly generated from
the late hadronic phase with very large radial flow
{\it i.e.} the value of $v_r$ is maximum for $M\sim m_\rho$ and lower 
at both sides of the $\rho$ mass. Resulting in the nonmonotonic
behaviour of the inverse slope, which contains
the effect of temperature ($T_{th}$) and radial flow ($v_r$)
as $T_{eff}=T_{th}+ M_{av}v_r^2/$. At large M ($>1.2$ GeV)  the flow
is predominantly longitudinal and hence the inverse slope
falls slowly with decreasing $M_{av}$.

In summary 
we have studied both the $M$ and $M_T$ 
distributions of dileptons from heavy ion collisions at LHC energy.
We have shown
that the effective temperature extracted from the inverse
slope of the $M_T$ spectra for various $M$
windows can be used as an efficient tool to characterize different
phases of the evolving matter.

\end{document}